\documentclass[aps,prl,reprint,superscriptaddress,amsmath,amssymb,longbibliography,floatfix]{revtex4-2}

\usepackage[T1]{fontenc}
\usepackage[utf8]{inputenc}
\usepackage{lmodern}
\usepackage{microtype}
\usepackage{bm,graphicx,xcolor,booktabs,amsmath}
\usepackage{hyperref}
\usepackage{braket}
\usepackage[Symbol]{upgreek}
\usepackage[percent]{overpic}
\usepackage{xcolor}

\hypersetup{colorlinks=true,linkcolor=blue!50!black,citecolor=blue!50!black,urlcolor=blue!50!black}
\graphicspath{{figures/}{figures/supplemental/}}

\newif\ifdraftnotes
\draftnotestrue

\newcommand{\figureasset}[4]{%
  \begin{figure}[t]
  \centering
  \IfFileExists{#1}{\includegraphics[width=\linewidth]{#1}}{%
    \fbox{\parbox[c][20mm][c]{0.91\linewidth}{\centering
      \textbf{#2}\\[5pt]\scriptsize\texttt{\detokenize{#1}}}}}
  \caption{#3}\label{#4}
  \end{figure}}

\newcommand{\widefigureasset}[4]{%
  \begin{figure*}[t]
  \centering
  \IfFileExists{#1}{\includegraphics[width=0.95\textwidth]{#1}}{%
    \fbox{\parbox[c][35mm][c]{0.91\textwidth}{\centering
      \textbf{#2}\\[5pt]\scriptsize\texttt{\detokenize{#1}}}}}
  \caption{#3}\label{#4}
  \end{figure*}}

\newcommand{\prlheading}[1]{\par\indent\textit{\textcolor{blue}{#1.}}---\ignorespaces}

\newcommand{\mA}{\mathcal{A}}

\newcommand{\rvec}{\mathbf{r}}
\newcommand{\kvec}{\mathbf{k}}
\newcommand{\vm}{\mathbf{m}}
\newcommand{\va}{\upalpha}
\newcommand{\vb}{\upbeta}

\newcommand{\dd}{\mathrm{d}}

\newcommand{\papertitle}{Collinear and Coplanar Altermagnetism in a Penrose Quasicrystal}
\newif\ifincludeendmatter
\includeendmatterfalse

\begin{document}
\title{\papertitle}

\author{Bai-He Yao}
\affiliation{School of Physics, Beijing Institute of Technology, Beijing 100081, China}
\author{Zhi-Yan Shao}
\email{3120225763@bit.edu.cn}
\affiliation{School of Physics, Beijing Institute of Technology, Beijing 100081, China}
\author{Ye Cao}
\email{ycao@bit.edu.cn}
\affiliation{School of Physics, Beijing Institute of Technology, Beijing 100081, China}
\author{Fan Yang}
\email{yangfan\_blg@bit.edu.cn}
\affiliation{School of Physics, Beijing Institute of Technology, Beijing 100081, China}

\date{\today}
\begin{abstract}

We generalize the previous group-theory based classification of collinear magnetism in the quasicrystal (QC) to general magnetic states. Taking the Penrose lattice as an example, we find that the magnetic states corresponding to 2D irreducible representations (IRRPs) of the $D_5$ point group represent for a new type of altermagnetism (AM). Unlike conventional AM wherein zero magnetism is protected by, e.g. rotation or mirror reflection, which globally flips the spin in the magnetic pattern, the zero magnetism in these 2D-IRRP states is instead protected by a special group-representation-orthogonality theorem which states that basis functions belonging to different IRRPs are mutually orthogonal. This special compensation mechanism and the absence of $\mathcal{PT}$ symmetry finally lead to AM states. Taking the Penrose-Hubbard model as an example, our linear-response theory and Ginzburg-Landau theory based analysis show that the largest portion of the parameter space is occupied by such 2D-IRRP states, which can be either collinear or coplanar AM states. These special AM states are characterized by spin split in the spin-resolved angular-resolved photoemission spectroscopy and the spin conductance with unique symmetry features. These intriguing collinear and coplanar AM states result from symmetry and can also emerge in other QCs.

\end{abstract}
\maketitle
\prlheading{Introduction}
In recent years, altermagnetism (AM), characterized by zero magnetism similar to antiferromagnetism (AFM) and spin-split band structure similar to ferromagnetism (FM)~\cite{mazinEditorialAltermagnetismANew2022,smejkalConventionalFerromagnetismAntiferromagnetism2022,smejkalEmergingResearchLandscape2022,baiAltermagnetismExploringNew2024,jungwirthAltermagnetismUnconventionalSpinordered2025a}, has attracted great research interest~\cite{hayamiMomentumDependentSpinSplitting2019,yuanGiantMomentumdependentSpin2020,mazinPredictionUnconventionalMagnetism2021,yuanPredictionLowZCollinear2021,smejkalCrystalTimereversalSymmetry2020,nakaSpinCurrentGeneration2019,maMultifunctionalAntiferromagneticMaterials2021,dornellasAltermagnetismCrystalSymmetry2026,zhouCrystalThermalTransport2024,urru$G$typeAntiferromagnetic$mathrmBiFeO_3$2025}. 
Researchers predict \cite{fenderAltermagnetismChemicalPerspective2025,leebSpontaneousFormationAltermagnetism2024,meierNetCompensatedAltermagnetism2026,vilaOrbitalspinLockingIts2025,ahnAntiferromagnetism$mathrmRuO_2$$d$wave2019,wuFermiLiquidInstabilities2007,wuDynamicGenerationSpinOrbit2004,qianFragileUnconventionalMagnetism2025,zhuDesignAltermagneticModels2025} and identify \cite{jiangMetallicRoomtemperatureDwave2025,osumiObservationGiantBand2024,reimersDirectObservationAltermagnetic2024,reichlovaObservationSpontaneousAnomalous2024,grahamLocalProbeEvidence2025,linObservationGiantSpin2024} possible AM materials, as well as explore possible applications of AM in spintronics \cite{hanElectrical180degSwitching2024,shaoSpinneutralCurrentsSpintronics2021,zhangPredictableGateFieldControl2024,zhouManipulationAltermagneticOrder2025}.

Symmetry classification \cite{brinkmanTheorySpinspaceGroups1966,liuSpinGroupSymmetryMagnetic2022,xiaoSpinSpaceGroups2024,chenEnumerationRepresentationTheory2024,jiangEnumerationSpinSpaceGroups2024,mcclartyLandauTheoryAltermagnetism2024} explains the difference between AM and other magnetism, especially AFM, in crystals. In an AFM state existing in an inversion ($\mathcal{P}$) symmetric system, the magnetism of two different parts connected by $\mathcal{P}$ is just different by a minus sign, leading to the combined $\mathcal{P}$ and time-reversal ($\mathcal{T}$) symmetry. Meanwhile, in the magnetic pattern of an AM state, some operations, such as mirror reflection or rotation, flip the spin while $\mathcal{P}$ does not, leading to zero magnetism and the absence of $\mathcal{PT}$ symmetry. These lead to the spin-split band structure, which can be detected by spin-resolved angular-resolved photoemission spectroscopy (sARPES) and spin conductance experiments.

Quasicrystal (QC) is a long-range ordered structure lacking translational symmetry \cite{levineQuasicrystalsNewClass1984,levineQuasicrystalsDefinitionStructure1986,socolarQuasicrystalsIIUnitcell1986}. 
This structural feature, combined with point group symmetry not allowed in crystals, leads to unique properties or phases \cite{andradeNonFermiLiquidBehaviorMetallic2015,araujoFragileMagneticOrder2024,kogaAntiferromagneticOrderHubbard2017,kogaFerrimagneticallyOrderedStates2022,caoKohnLuttingerMechanismDriven2020,huangQuantumSpinHall2018,krausTopologicalStatesAdiabatic2012,liuNematicSuperconductivityIts2024,uriSuperconductivityStrongInteractions2023,wesselQuantumAntiferromagnetismQuasicrystals2003}. 
In the absence of translational symmetry, there can be no more than one symmetry center in the QC. 
If a magnetic pattern is not $\mathcal{P}$-odd about the unique symmetry center while spin flips under some point-group operation, the corresponding state is AM. 
Such a routine makes it possible to search for AM in QCs.

Progress has been made in the search of AM states in QCs~\cite{chenAltermagnetismQuasicrystals2026,liUnconventionalAltermagnetismQuasicrystals2026,shaoClassificationMagnetismAltermagnetism2026,jiNeelOrderedMagnetic2025}. 
For QCs without spin-orbit coupling (SOC), AM and other magnetism like AFM and FM or ferrimagnetism (FIM) can be classified by irreducible representations (IRRPs) of point group due to the lack of translational symmetry \cite{shaoClassificationMagnetismAltermagnetism2026}. Specially, the magnetism of N\'eel states in bipartite QCs can also be judged by lattice geometry \cite{jiNeelOrderedMagnetic2025}. 
However, the classification \cite{shaoClassificationMagnetismAltermagnetism2026} only focuses on the magnetic states corresponding to 1D IRRPs, which represent collinear magnetic states. 
Magnetic states corresponding to higher dimensional IRRPs in QCs, which are possibly non-collinear orders, have not been studied yet.
Will these states lead to novel types of AM?

In this paper, we classify the magnetic states on a Penrose-Hubbard model based on the IRRPs of the $D_5$ point group, and find that all non-identity IRRPs correspond to AM. Especially, we identify AM states corresponding to 2D IRRPs whose zero net magnetism is protected by a special group-representation-orthogonality theorem which states that basis functions belonging to different IRRPs are mutually orthogonal. Such a special compensation mechanism for zero magnetism distinguishes these 2D-IRRP AM states from conventional AM states in which zero magnetism is protected by such simple symmetry operations as rotation or mirror reflection which globally flip the spin pattern. Our linear-response theory combined with Ginzburg-Landau (G-L) theory based analysis reveal that the largest portion of the parameter space is occupied by these 2D-IRRP states, which take the form of collinear or coplanar AM states. These AM states are characterized by spin split in the sARPES and spin conductance with unique symmetry features. The central point here results from symmetry and applies to other QC systems.

\begin{figure}[t]
  \centering
  \begin{minipage}[c]{0.46\columnwidth}
    \centering
    \begin{overpic}[width=\linewidth]        
      {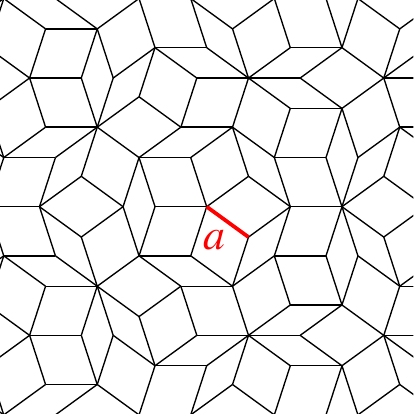}
      \put(2,96){(a)}
    \end{overpic}
  \end{minipage}
  \hfill
  \begin{minipage}[c]{0.52\columnwidth}
    \centering
    \begin{overpic}[width=\linewidth]
       {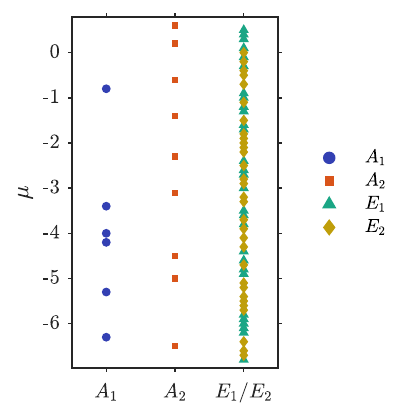}
       \put(2,90){(b)}
    \end{overpic}
      
  \end{minipage}

  \caption{
  (a) A view of the Penrose tiling, with the rhombus edge length $a$ highlighted in red.
  (b) Symmetry classes of the leading
  susceptibility eigenvector as a function of the
  chemical potential $\mu$. The horizontal axis distinguishes the 1D IRRPs $A_1$ and $A_2$ from the 2D IRRPs $E_1$ and $E_2$.
  }
  \label{fig:leading-irrep}
\end{figure}

\prlheading{Model and approach}
We study magnetic instabilities of the Hubbard model on a $D_5$-symmetric Penrose lattice with open boundaries and no spin--orbit coupling:
\begin{equation}
\begin{aligned}
H&=-\sum_{ij,\sigma}t_{ij}c^{\dagger}_{i\sigma}c_{j\sigma}
-\mu\sum_{i,\sigma}n_{i\sigma}
+U\sum_i n_{i\uparrow}n_{i\downarrow},\\
t_{ij}&=t_0e^{-r_{ij}/a}.
\end{aligned}
\label{eq:hamiltonian}
\end{equation}
Here $c^{\dagger}_{i\sigma}$ creates an electron with spin $\sigma$ at site $i$, $n_{i\sigma}=c^{\dagger}_{i\sigma}c_{i\sigma}$, and $r_{ij}=|\rvec_i-\rvec_j|$. The hopping decay length $a$ equals the rhombus edge length highlighted in Fig.~\ref{fig:leading-irrep}(a), and $t_0=1$ sets the energy unit. The chemical potential $\mu$ adjusts the filling and $U>0$ is the on-site repulsion.

We identify magnetic instabilities from the static spin susceptibility ~\cite{shaoClassificationMagnetismAltermagnetism2026}. With $S_i^z=(n_{i\uparrow}-n_{i\downarrow})/2$, we define
\begin{equation}
\chi_{ij}=\int_0^{\beta}\dd\tau\,
\langle T_\tau S_i^z(\tau)S_j^z(0)\rangle,
\end{equation}
where $\beta$ is the inverse temperature, and we take $\beta\rightarrow\infty$. Here, $\chi_{ij}$ describes the response of the magnetic moment at site $i$ to the self-consistent exchange field at site $j$. When $U$ enhances above its critical value $U_c$, $\chi$ diverges, signaling the onset of magnetic order. The resultant magnetic pattern is approximated as the eigenvector $\boldsymbol{\xi}\equiv(\xi_{i})$ corresponding to the largest eigenvalue $\lambda$ of the matrix $\chi\equiv(\chi_{ij})$ for $U\to U_c^{-}$ through the eigenvalue problem~\cite{shaoClassificationMagnetismAltermagnetism2026}  
\begin{equation}\label{eigen_function}
\sum_j\chi_{ij}\xi_{j}=\lambda\xi_{i}.
\end{equation}

When the leading eigenvector $\boldsymbol{\xi}$ solved for Eq. (\ref{eigen_function}) is nondegenerate, it represents for collinear magnetic state wherein the magnetic moments on all sites $\{\vm_i\}$ polarize along a common spin axis, arbitrarily chosen as spin-$z$ axis, and $\xi_{i}$ is proportional to the algebraic value of $\vm_i$ along spin-$z$ axis. In contrast, when two degenerate leading eigenvectors $\boldsymbol{\xi}^{(a)} (a=1,2)$ emerge, the realized magnetic pattern is the vector superposition of two basis patterns with each hosting its own polarization direction, with the relative orientation of the two polarization directions determined by energy minimization through a G-L theory based analysis.  Consequently, when the two polarization directions are parallel or antiparallel, the realized state is collinear state; otherwise it is coplanar one. See the Supplementary Materials (SM) \cite{SM} for details.

\prlheading{Magnetic classification}
Since $\chi$ is invariant under $D_5$, its eigenvectors corresponding to the same eigenvalue form an IRRP of $D_5$. Based on this, we classify the magnetic phases according to these IRRPs summarized in Table~\ref{tab:d5-irreps}. The columns list the IRRPs of $D_5$, their generators' representation matrices and polynomial basis functions, the mechanisms enforcing magnetic compensation, and the magnetic classification of the states.
\begin{table*}[t]
\centering
\small
\setlength{\tabcolsep}{9pt}
\renewcommand{\arraystretch}{1.6}
\caption{IRRPs of $D_5$, their magnetic classification and compensation mechanism. Here $\theta=2\pi/5$, the $C_5$ rotation and mirror-reflection $\sigma_x$ are generators of $D_5$. The polynomial basis functions only illustrate the transformation properties.}
\label{tab:d5-irreps}
\begin{tabular}{cccccc}
\toprule
IRRP
& $D(C_5)$
& $D(\sigma_x)$
& Basis functions
& Compensation Mechanism
& Classification \\
\midrule

$A_1$
& $1$
& $1$
& $1$
& Not enforced
& FIM \\

$A_2$
& $1$
& $-1$
& $\operatorname{Re}(x+iy)^5*\operatorname{Im}(x+iy)^5$
& Mirror-odd symmetry
& AM \\

\midrule

$E_1$
&
$\begin{pmatrix}
\cos\theta & -\sin\theta \\
\sin\theta & \cos\theta
\end{pmatrix}$
&
$\begin{pmatrix}
1 & 0 \\
0 & -1
\end{pmatrix}$
&
$(x,y)$
& group-representation-orthogonality
& AM \\[8pt]

$E_2$
&
$\begin{pmatrix}
\cos 2\theta & -\sin 2\theta \\
\sin 2\theta & \cos 2\theta
\end{pmatrix}$
&
$\begin{pmatrix}
1 & 0 \\
0 & -1
\end{pmatrix}$
&
$(x^2-y^2,\,2xy)$
& group-representation-orthogonality
& AM \\[4pt]

\bottomrule
\end{tabular}
\end{table*}

The two 1D IRRPs $A_1$ and $A_2$ have been studied in Ref.\cite{shaoClassificationMagnetismAltermagnetism2026}: For the identity IRRP $A_1$, its basis function is invariant under full $D_5$. As there is no symmetry operation to ensure zero magnetism, this IRRP corresponds to FIM. For the other 1D IRRP $A_2$, the mirror reflection $\sigma_x$ flips the spin in the magnetic pattern, which ensures zero magnetism. Further more, as the Penrose lattice lacks inversion symmetry, the magnetic states cannot be $\mathcal{PT}$-invariant. Thus, $A_2$ corresponds to AM. 

The two 2D IRRPs $E_1$ and $E_2$ are new. Within either $E_1$ or $E_2$, the two basis patterns $\{\xi^{(1)}_i\}$ and $\{\xi^{(2)}_i\}$ can be chosen to be even and odd under the mirror reflection $\sigma_x$ respectively. The two bases are mixed under the $C^1_5$ rotation through the mixing matrix listed in Table~\ref{tab:d5-irreps}. The realized magnetic pattern $\{\vm_i\}$ is described by the following vector sum of $\{\xi^{(1)}_i\}$ and $\{\xi^{(2)}_i\}$,
\begin{equation}
\vm_i=\va\xi_{i}^{(1)}+\vb\xi_{i}^{(2)}.
\label{eq:mixed-mode}
\end{equation}
Here $\va$ and $\vb$ are the polarization vectors of the two magnetic patterns. The resultant magnetic state is collinear or coplanar, determined by whether $\va\parallel\vb$ or not .

It's interesting to note that, for the 2D-IRRP state $\{\vm_i\}$ described by Eq.~(\ref{eq:mixed-mode}), zero magnetism can always be protected by the compensation mechanism through the group-representation-orthogonality theorem: The basis functions belonging to different IRRPs of a given group are mutually orthogonal. To show this, let us choose the uniform pattern $\boldsymbol{u}=1/\sqrt{N}\{1,1,\cdots,1\}$, which belongs to the $A_1$ IRRP. Since the two eigen patterns $\boldsymbol{\xi}^{(a)}(a=1,2)$ belong to a different IRRP, i.e. $E_{1}$ or $E_{2}$, we have 
\begin{equation}
  \braket{\boldsymbol{u}|\boldsymbol{\xi}^{(a)}}
  =1/\sqrt{N}\sum_i\xi_{i}^{(a)}
  =0,
  \qquad a=1,2.
  \label{eq:irrep-orthogonality-compensation}
\end{equation}
Therefore, zero magnetism is guaranteed in both $\{\xi^{(1)}_i\}$ and $\{\xi^{(2)}_i\}$ and hence their vector superposition $\{\vm_i\}$.   

Due to exact magnetic compensation and lack of $\mathcal{PT}$ symmetry, these 2D-IRRP states are AM. These intriguing AM states are distinguished from traditional AM because the special group-representation-orthogonality protected compensation mechanism is different from the traditional compensation mechanism protected by such simple geometry operations as rotation and mirror reflection which globally flip the spin pattern.


\prlheading{Collinear and Coplanar AM belonging to 2D IRRPs} Figure~\ref{fig:leading-irrep}(b) shows the leading IRRPs for $\mu$ ranging from $-6.8t_0$ to $0.6t_0$ in steps of $0.1t_0$. The Penrose tilting adopted hosts $3031$ sites with a radius of $28a$. Obviously, the 2D IRRPs $E_1$ and $E_2$ occupy the largest portion of the parameter space. Note that the phases vary irregularly with $\mu$, possibly reflecting the fractal properties of the density of states in Penrose quasicrystals~\cite{vignoloEnergyLandscapePenrose2016}.

\begin{figure}[t]
  \centering
  \begin{minipage}[t]{0.45\columnwidth}
    \centering
    \begin{overpic}[width=\linewidth]{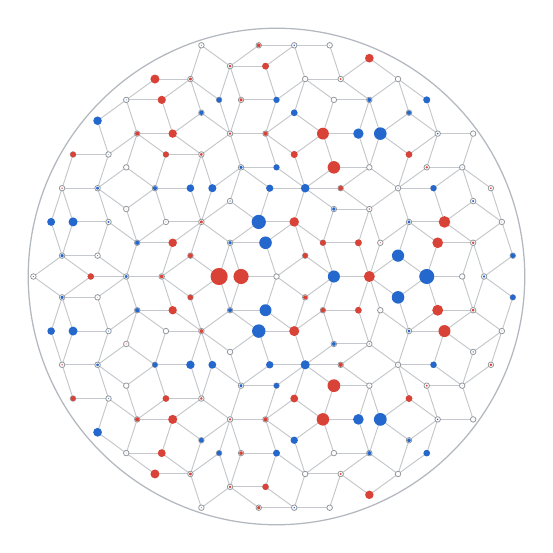}
    \put(3,94){\small\bfseries (a)}
  \end{overpic}
  \label{fig2a}
  \end{minipage}
  \hfill
  \begin{minipage}[t]{0.45\columnwidth}
    \centering
    \begin{overpic}[width=\linewidth]{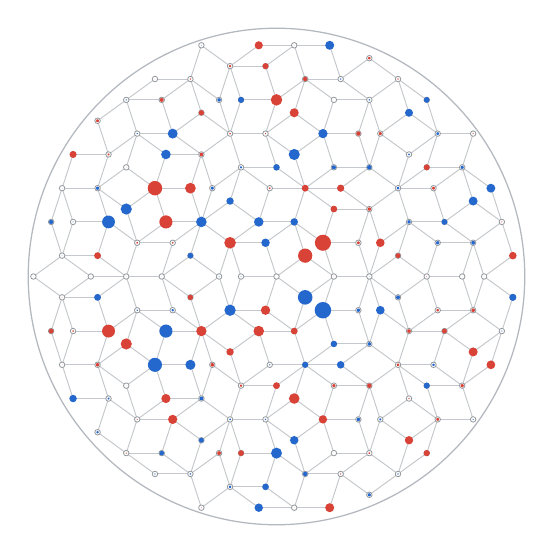}
    \put(3,94){\small\bfseries (b)}
  \end{overpic}
  \label{fig2b}
  \end{minipage}

  \vspace{-1.2em}

  \begin{minipage}[t]{0.48\columnwidth}
    \centering
    \begin{overpic}[width=\linewidth]{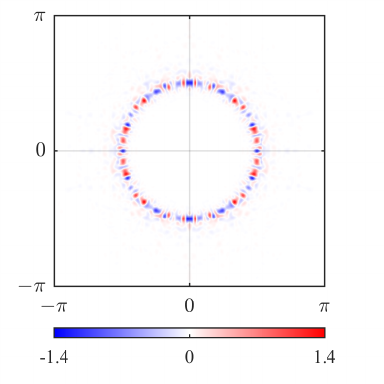}
    \put(14,89){\small\bfseries (c) $\mA_\uparrow(\kvec)-\mA_\downarrow(\kvec)$}
  \end{overpic}
  \end{minipage}
  \hfill
  \begin{minipage}[t]{0.48\columnwidth}
    \centering
    \begin{overpic}[width=\linewidth]{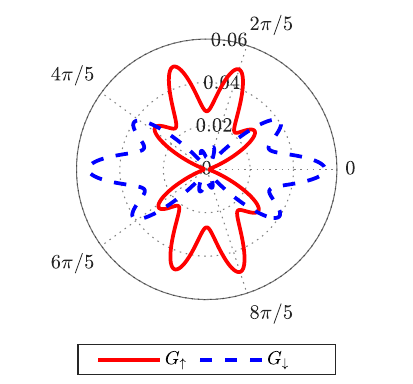}
    \put(3,89){\small\bfseries (d) }
  \end{overpic}
  \end{minipage}
  \caption{Collinear magnetic order belonging to the $E_2$ IRRP and its observable signatures at $\mu=-0.4$. The observables reflect the symmetry of the magnetic order. (a,b) Real-space patterns A and B, respectively, cropped to a central disk of radius $7a$ from the full radius-$28a$ Penrose tilting. (c) The sARPES of pattern A at $\omega=-0.05t_0$. (d) Spin conductance of pattern A. Panels (c,d) show the spin split feature. }
  \label{fig:collinear_observables}
\end{figure}
\begin{figure*}[t]
  \centering

  \begin{minipage}[t]{0.24\textwidth}
    \vspace{0pt}
    \centering
    \begin{overpic}[width=\linewidth]{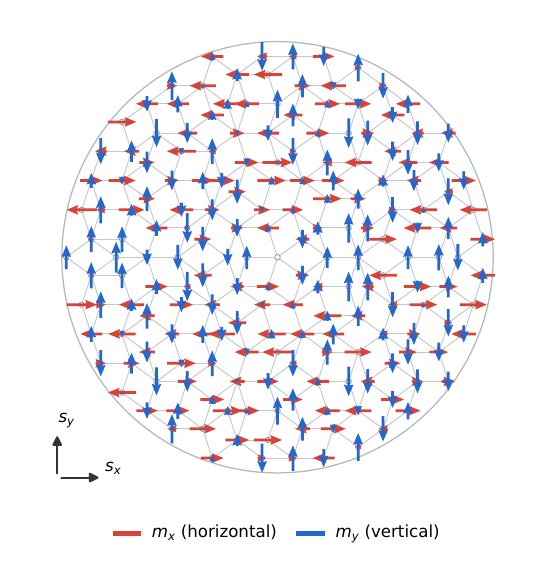}
      \put(3,90){\small\bfseries (a)}
    \end{overpic}
  \end{minipage}%
  \hfill
  \begin{minipage}[t]{0.74\textwidth}
    \vspace{0pt}
    \centering
    \begin{overpic}[width=\linewidth]{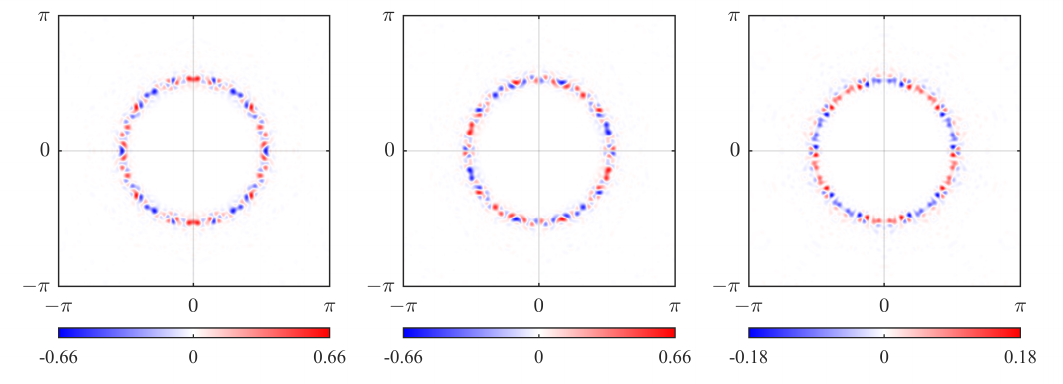}
      \put(6,32){\small
        \textbf{(c)} $\mA_{x+}-\mA_{x-}$}
      \put(38.5,32){\small
        \textbf{(d)} $\mA_{y+}-\mA_{y-}$}
      \put(71,32){\small
        \textbf{(e)} $\mA_{z+}-\mA_{z-}$}
    \end{overpic}
  \end{minipage}

  \par\vspace{-0.8em}

  \begin{minipage}[t]{0.24\textwidth}
    \vspace{0pt}
    \centering
    \begin{overpic}[width=\linewidth]{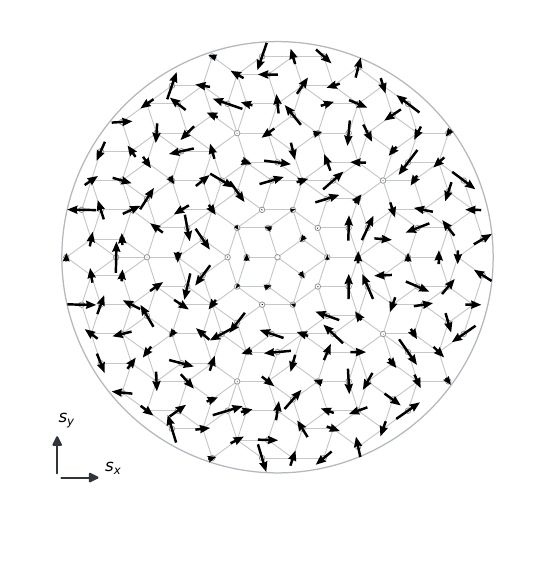}
      \put(3,90){\small\bfseries (b)}
    \end{overpic}
  \end{minipage}%
  \hfill
  \begin{minipage}[t]{0.74\textwidth}
    \vspace{0pt}
    \centering
    \begin{overpic}[width=\linewidth]{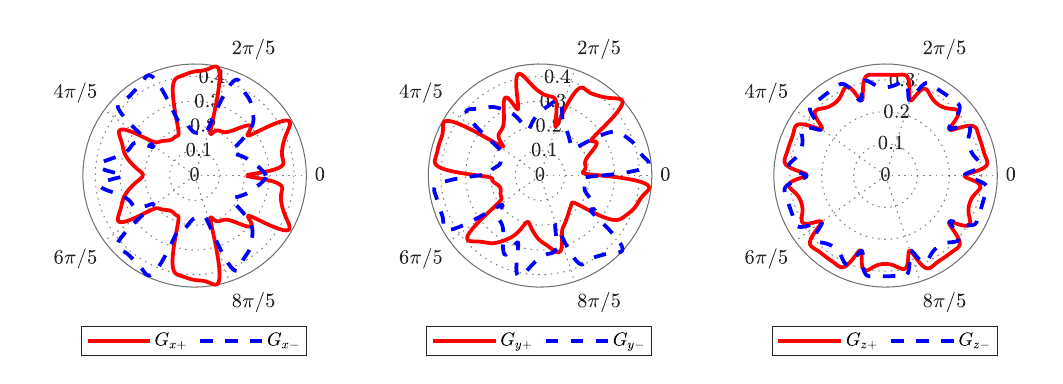}
      \put(3,32){\small \textbf{(f)} $x$}
      \put(35.5,32){\small \textbf{(g)} $y$}
      \put(68,32){\small \textbf{(h)} $z$}
    \end{overpic}
  \end{minipage}

  \caption{
    Coplanar magnetic order in the $E_1$ IRRP and its signatures for $\mu=-0.3$.
(a) Mirror-even $\boldsymbol{\xi}^{(1)}$ (red, polarized along spin-$x$) and mirror-odd $\boldsymbol{\xi}^{(2)}$ (blue, polarized along spin-$y$) patterns.
(b) Coplanar pattern C obtained from vector sum of $\boldsymbol{\xi}^{(1)}$ and $\boldsymbol{\xi}^{(2)}$, with arrow direction and size indicating spin orientation and moment magnitude, respectively.
Panels (a,b) use the same central crop as Fig.~\ref{fig:collinear_observables}(a,b).
     The $x$-, $y$-, and $z$- resolved (c--e) sARPES for $\omega=-0.05t_0$, and (f-h) angle-dependent conductance, respectively. In (f-h), red solid and blue dashed curves denote opposite spin projections. With the mirror line along the $x$-axis, the spin split is mirror-even in (c, f), mirror-odd in (d, g), and $C_5$-invariant in (e, h).}
  \label{fig:coplanar_observables}
\end{figure*}

For the 2D IRRPs $E_{1,2}$, a G-L theory based analysis provided in the SM \cite{SM} suggests two possibilities for the relation between the polarization vectors $\va$ and $\vb$ in Eq. (\ref{eq:mixed-mode}): (a) $\va\parallel\vb$, leading to collinear AM state. Furthermore, we have $\vb=$tan$(k\pi/10)\va$ with $k=0,\ldots,9$. However, through $C_5$ rotation or time-reversal operation, these 10 possibilities can be reduced to two symmetry-inequivalent ones $(\va,0)$ and $(0,\vb)$. This gives rise to two symmetry-inequivalent patterns $\boldsymbol{\xi}^{(1)}$ and $\boldsymbol{\xi}^{(2)}$. (b) $\va\perp\vb,|\va|=|\vb|$, leading to coplanar AM state. We compare the mean-field energies at $1000$ sampled points in the $(\mu,U)$ parameter space associated with 2D IRRPs. Our mean-field calculations suggest that a portion of 99.3\% of the parameter space favoring 2D IRRPs satisfies (b), suggesting that the coplanar AM states dominate the parameter space.

These obtained collinear and coplanar AM states can be characterized by the sARPES and spin conductance~\cite{shaoClassificationMagnetismAltermagnetism2026,chenAltermagnetismQuasicrystals2026,liUnconventionalAltermagnetismQuasicrystals2026}. Concretely, we calculate the projections of the ARPES weight or the conductance onto oppositely spin-polarized states along $x,y,z$ axes, dubbed as $x$-, $y$-, $z$- resolved, respectively. Specifically, the ``positively'' and ``negatively'' spin-polarized states along $x$ and $y$ axes are $\ket{\pm}_x=(\ket{\uparrow}\pm\ket{\downarrow})/\sqrt{2}$
and $\ket{\pm}_y=(\ket{\uparrow}\pm i\ket{\downarrow})/\sqrt{2}$ in the spin-$z$ representation. Here, $z$ is chosen along the polarization direction for collinear states and normal to the spin plane for coplanar states. See the SM \cite{SM}.

For the collinear states, the two symmetry-inequivalent patterns $\boldsymbol{\xi}^{(1)}$ and $\boldsymbol{\xi}^{(2)}$ are labeled as A and B, as shown in Figs.~\ref{fig:collinear_observables}(a) and (b) respectively. Obviously, A and B are even and odd under the mirror reflection $\sigma_x$. The compensation mechanism for B is the same as that for conventional AM state such as the $A_2$-IRRP state, because $\sigma_x$ globally flips the spin pattern in B and protects zero magnetism. The situation for A is distinct: In A, the spin pattern keeps invariant under $\sigma_x$, and no other point-group operation globally flips the spin pattern. However, the special group-representation-orthogonality theorem ensures zero magnetism for pattern A. This distinguishes the collinear AM states described by pattern A from conventional collinear AM states. Figs.~\ref{fig:collinear_observables}(c, d) show clear spin split in the sARPES and spin conductance of pattern A, verifying its AM character. Furthermore, both calculated experimental figures are mirror-even: Reflection $\sigma_x$ preserves each spin component, rather than exchanging opposite-spin responses as in conventional mirror protected AM.

For the coplanar states, the mirror-even pattern $\boldsymbol{\xi}^{(1)}$ and mirror-odd pattern $\boldsymbol{\xi}^{(2)}$ are mixed with their associate spin-polarization vectors $\va$ and $\vb$ mutually orthogonal. In Fig.~\ref{fig:coplanar_observables}(a), the red and blue arrows represent the magnetic patterns contributed from $\{\xi^{(1)}_i\va\}$ and $\{\xi^{(2)}_i\vb\}$ with $\va$ and $\vb$ lying on the spin-$x$ and spin-$y$ axes, respectively.  
Fig.~\ref{fig:coplanar_observables}(b) shows the magnetic pattern C obtained from vector sum of $\{\xi^{(1)}_i\va\}$ and $\{\xi^{(2)}_i\vb\}$ through Eq.~(\ref{eq:mixed-mode}), forming a coplanar AM state. In this coplanar state, group-representation-orthogonality theorem ensure magnetic compensation, distinguishing it from conventional AM states. Besides, for this equal-amplitude coplanar state, the magnetic compensation is also enforced by the combined spatial and spin rotations, e.g. spin-space rotation and real-space rotation with $2\pi/5$: $U_z(2\pi/5)R_z(2\pi/5)$.

Figs.~\ref{fig:coplanar_observables}(c-e) and Figs.~\ref{fig:coplanar_observables}(f-h) show the sARPES and spin conductance of the obtained coplanar magnetic state respectively, which clearly exhibit spin splitting, verifying the AM character. The $x$-resolved and $y$-resolved sARPES or spin conductance are mirror-even and odd respectively, reflecting the symmetry of the $\boldsymbol{\xi}^{(1)}$ and $\boldsymbol{\xi}^{(2)}$ components. Again, the mirror-even $x$-resolved sARPES or spin conductance is distinguished from conventional AM state. Besides, the $z$-resolved sARPES or spin conductance in (e) or (h) is mirror-odd under $\sigma_x$ and is $C_5$-invariant, with the $C_5$-invariance reflecting the symmetry under the combination of real-space rotation $R_z(2\pi/5)$ and spin-space rotation $U_z(2\pi/5)$ detailed in the SM \cite{SM}. This symmetry difference of sARPES and spin conductance among different spin-polarization directions helps us to recognize the spin-$xy$ plane. 


\prlheading{Discussion and conclusion}
By symmetry analysis, we point out that all magnetic states corresponding to non-identity IRRPs of $D_5$ point group are AM. 
Particularly, we find that the 2D-IRRP states, either collinear or coplanar, represent for a new type of AM states in which the zero magnetism is protected by the group-representation-orthogonality. This new compensation mechanism distinguishes these 2D-IRRP AM states from conventional AM states in which the zero magnetism is protected by simple geometry operation which globally flips the spin pattern. These AM states are characterized by the spin split in the sARPES and spin conductance with unique symmetry features.     

Note that the group-representation-orthogonality protected compensation mechanism proposed here is based on linear-response theory, which only works well for weak magnetic order obtained at $U$ slightly stronger than $U_c$. For large $U$, the magnetic state might be a mixture of different IRRPs~\cite{baumExoticElectronicOrder2026}. In such cases, the collinear pattern A shown in Fig.~2(a) might only be an approximate AM. However, the coplanar pattern C shown in Fig.~3(b) is still exact AM protected by rotation symmetry. 

The central point of this work is general since it results from symmetry analysis and does not depend on any specific model. 
Similar classification can be extended to magnetic states corresponding to 2D IRRPs of $D_n$ group for general $n$.
For odd $n$, it is predicted that the magnetic states corresponding to all 2D IRRPs are AM. For even $n$, it is predicted that the magnetic states corresponding to 2D IRRP $E_m$ ($m=1,\cdots,\frac{n}{2}-1$ representing the angular momentum of the basis function) are AFM for odd $m$ while AM for even $m$.



\prlheading{Note Added} On finishing this work, we notice that a preprint work~\cite{hongAltermagneticGroundStates2026} also obtained coplanar altermagnetic order in a classical Heisenberg spin model on the Ammann--Beenker tiling. Our work is independent of that work, and is distinguished from that work in the following aspects: (1) interaction-driven magnetic instabilities in a Penrose Hubbard model, (2) magnetic compensation enforced by group-representation-orthogonality for arbitrary vector combinations within a non-identity doublet, (3) momentum-resolved spin splitting accessible to sARPES, distinct mirror and fivefold rotational properties of the sARPES.



\prlheading{Acknowledgments}F.Y. is supported by the National Natural Science Foundation of China under the Grant Nos. 12574141, 12234016.

\prlheading{Data availability}
All data are displayed in the main text and the supplementary material. All data are available from the corresponding author upon request.

\nocite{shaoClassificationMagnetismAltermagnetism2026}

\nocite{liuSpinGroupSymmetryMagnetic2022}
\bibliographystyle{apsrev4-2}
\bibliography{refs/ref}

\clearpage
\onecolumngrid
\setcounter{section}{0}
\setcounter{subsection}{0}
\setcounter{subsubsection}{0}
\setcounter{equation}{0}
\setcounter{figure}{0}
\setcounter{table}{0}
\setcounter{secnumdepth}{1}
\setcounter{tocdepth}{1}
\renewcommand{\thesection}{\Roman{section}}
\renewcommand{\theequation}{S\arabic{equation}}
\renewcommand{\thefigure}{S\arabic{figure}}
\renewcommand{\thetable}{S\arabic{table}}

\renewcommand{\theHsection}{SM.\arabic{section}}
\renewcommand{\theHsubsection}{\theHsection.\arabic{subsection}}
\renewcommand{\theHsubsubsection}{\theHsubsection.\arabic{subsubsection}}
\renewcommand{\theHequation}{SM.\arabic{equation}}
\renewcommand{\theHfigure}{SM.\arabic{figure}}
\renewcommand{\theHtable}{SM.\arabic{table}}

\pdfbookmark[0]{Supplemental Material}{supplemental-material}
\begin{center}
  \large\bfseries
  Supplemental Material for ``\papertitle''
\end{center}
\medskip
\begin{center}
  \textbf{CONTENTS}
\end{center}
\begingroup
\small
\renewcommand{\arraystretch}{1.4}
\noindent\begin{tabular}{@{}p{0.90\linewidth}@{\hspace{1em}}r@{}}
\hyperref[sus:section]{\ref*{sus:section}. Static susceptibility and magnetic ordering modes}
  & \pageref{sus:section} \\
\hyperref[app:gl-selection]{\ref*{app:gl-selection}. Ginzburg--Landau symmetry analysis of 2D irreducible representations}
  & \pageref{app:gl-selection} \\
\hyperref[app:mf_energy]{\ref*{app:mf_energy}. Mean-field energy comparison}
  & \pageref{app:mf_energy} \\
\hyperref[sp:section]{\ref*{sp:section}. Symmetry of the spin-resolved spectrum}
  & \pageref{sp:section} \\
\hyperref[sm:more_results]{\ref*{sm:more_results}. Additional results for two-dimensional IRRPs}
  & \pageref{sm:more_results}
\end{tabular}
\par\endgroup
\medskip

\section{Static susceptibility and magnetic ordering modes}
\label{sus:section}

Our analysis of magnetic ordering modes builds on the point-group approach to magnetism in quasicrystals developed in Ref.~\cite{shaoClassificationMagnetismAltermagnetism2026}.
We start from the Hubbard Hamiltonian in a weak static field as in 
\begin{equation}
 \begin{aligned}
 H_h&=H-\sum_i h_i S_i^z,\\
 H&=H_{\mathrm{TB}}+U\sum_i n_{i\uparrow}n_{i\downarrow},\\
 H_{\mathrm{TB}}&=-\sum_{ij,\sigma}t_{ij}c_{i\sigma}^\dagger c_{j\sigma}
                    -\mu\sum_{i,\sigma}n_{i\sigma},
 \end{aligned}
 \label{sus:hamiltonian}
\end{equation}
where $t_{ij}=t_0e^{-r_{ij}/a}$ and $S_i^z=(n_{i\uparrow}-n_{i\downarrow})/2$. $H_{\mathrm{TB}}$ is the tight-binding Hamiltonian.
For a nondegenerate paramagnetic ground state $\ket{0}$, first-order perturbation theory gives $\langle S_i^z\rangle_h=\sum_j\chi^{zz}_{ij}h_j+O(h^2)$, with
\begin{equation}
 \chi_{ij}=2\operatorname{Re}\sum_{r\ne0}
 \frac{\bra{0}S_i^z\ket{r}\bra{r}S_j^z\ket{0}}{E_r-E_0}.
 \label{sus:lehmann}
\end{equation}
Here $\ket{r}$ and $E_r$ are the many-body eigenstates and energies of $H$.
The corresponding finite-temperature static susceptibility is
\begin{equation}
 \chi_{ij}=\int_0^\beta d\tau\,
 \langle T_\tau\,\delta S_i^z(\tau)\delta S_j^z(0)\rangle,
 \qquad \beta=1/T,
 \label{sus:definition}
\end{equation}
where $\delta S_i^z=S_i^z-\langle S_i^z\rangle$ and $k_B=\hbar=1$.

To match the normalization used in our numerical calculations, we define the bare susceptibility as $\chi^0_{ij}\equiv2\chi^{zz,0}_{ij}$, where $\chi^{zz,0}$ is evaluated using $H_{\mathrm{TB}}$.
Let the hopping eigenstates satisfy $\sum_j(-t_{ij})\phi_{jn}=\varepsilon_n\phi_{in}$ and write $c_{i\sigma}=\sum_n\phi_{in}d_{n\sigma}$, so that $H_{\mathrm{TB}}=\sum_{n\sigma}(\varepsilon_n-\mu)d_{n\sigma}^\dagger d_{n\sigma}$.
Wick contraction and integration over imaginary time then give
\begin{equation}
 \chi^0_{ij}=\sum_{mn}
 \phi_{im}^*\phi_{in}\phi_{jn}^*\phi_{jm}
 \frac{f(\varepsilon_m-\mu)-f(\varepsilon_n-\mu)}
      {\varepsilon_n-\varepsilon_m},
 \label{sus:static-lindhard}
\end{equation}
where $f(e)=(e^{\beta e}+1)^{-1}$ is the Fermi--Dirac distribution.
For equal energies, including distinct degenerate states, the quotient is understood as $-f'(\varepsilon_m-\mu)$.
For real hopping, the $\phi_{in}$ can be chosen real, and $\chi^0$ is real symmetric.

For uniform on-site $U$, the random phase approximation (RPA) susceptibility in the same normalization satisfies
\begin{equation}
 \begin{aligned}
 \boldsymbol\chi
 &=\boldsymbol\chi^0+\boldsymbol\chi^0U\boldsymbol\chi\\
 &=[\mathbb I-\boldsymbol\chi^0U]^{-1}\boldsymbol\chi^0.
 \end{aligned}
 \label{sus:dyson}
\end{equation}
Uniform $U$ preserves the eigenvectors of $\chi^0$ and maps each eigenvalue $\lambda_n^0$ to $\lambda_n^0/(1-U\lambda_n^0)$.
Their ordering is therefore unchanged for $0< U<U_c$, with
\begin{equation}
 U_c=\frac{1}{\lambda_{\max}^0},
 \label{sus:critical-u}
\end{equation}
where $\lambda_{\max}^0$ is the largest eigenvalue of $\chi^0$.
We therefore extract the spatial ordering modes from its leading eigenspace; their collinear or coplanar combination is determined by the Ginzburg--Landau (G--L) analysis and mean-field energy comparison.
\section{Ginzburg--Landau symmetry analysis of 2D irreducible representations}
\label{app:gl-selection}

For an isolated twofold-degenerate $E_\ell$ doublet ($\ell=1,2$), choose real modes satisfying $\sum_i\xi_i^{(a)}\xi_i^{(b)}=\delta_{ab}$ and write
\begin{equation}
 \begin{aligned}
 \mathbf m_i
 &=\va\xi_i^{(1)}
   +\vb\xi_i^{(2)},\\
 \xi_i&=\xi_i^{(1)}+i\xi_i^{(2)},\qquad
 \boldsymbol\psi=\tfrac12(\va-i\vb).
 \end{aligned}
 \label{app:eq:gl-order}
\end{equation}
The magnetic texture is then
\begin{equation}
 \mathbf m_i=\boldsymbol\psi\xi_i+\boldsymbol\psi^*\xi_i^*.
 \label{app:eq:gl-real-texture}
\end{equation}
We adopt the convention in which the complex mode transforms as $\xi_i\mapsto e^{-2\pi i\ell/5}\xi_i$ under $C_5$, with the corresponding coefficient transformation $\boldsymbol\psi\mapsto e^{2\pi i\ell/5}\boldsymbol\psi$.
A reference mirror sends $\boldsymbol\psi\mapsto\boldsymbol\psi^*$, while time reversal sends $\boldsymbol\psi\mapsto-\boldsymbol\psi$.
Define the spin scalars $s=\boldsymbol\psi^*\!\cdot\boldsymbol\psi$ and $q=\boldsymbol\psi\cdot\boldsymbol\psi$.
In the absence of spin--orbit coupling, spin-rotation symmetry and $D_5$ symmetry give the general uniform free-energy expansion through quartic order,
\begin{equation}
 F-F_0=r s+u s^2+v|q|^2+O(|\boldsymbol\psi|^6).
 \label{app:eq:gl-fourth}
\end{equation}
The invariants can be expressed as
\begin{align}
 s&=\tfrac14(|\va|^2+|\vb|^2),\nonumber\\
 |q|^2&=\tfrac1{16}\bigl[(|\va|^2-|\vb|^2)^2
                  +4(\va\cdot\vb)^2\bigr].
 \label{app:eq:gl-invariants}
\end{align}
Since $0\leq|q|^2\leq s^2$, $v>0$ selects $q=0$ at fixed $s$, requiring $|\va|=|\vb|$ and $\va\cdot\vb=0$.
This defines the coplanar state C.
For $v<0$, the minimum has $|q|^2=s^2$, requiring linearly dependent $\va$ and $\vb$ and hence a collinear texture.
Near a continuous instability, these alternatives assume a stable quartic expansion, $u>0$ and $u+v>0$; the case $v=0$ requires higher-order terms.

A general collinear state can be parametrized as
\begin{equation}
 \begin{aligned}
 \boldsymbol\psi&=\frac{\rho}{2}e^{-i\varphi}\hat{\mathbf n},\\
 \mathbf m_i(\varphi)
 &=\rho\left[\cos\varphi\,\xi_i^{(1)}
            +\sin\varphi\,\xi_i^{(2)}\right]\hat{\mathbf n},
 \end{aligned}
 \label{app:eq:gl-collinear-angle}
\end{equation}
where $\hat{\mathbf n}$ is a unit vector in spin space and $\varphi$ specifies mixing in the mode subspace.
The quartic free energy leaves $\varphi$ undetermined.
Since $q$ transforms as $q\mapsto e^{4\pi i\ell/5}q$, the leading angular anisotropy for either doublet occurs at tenth order:
\begin{equation}
 F_{10}^{\mathrm{an}}
 =w(q^5+q^{*5})
 =\frac{w\rho^{10}}{2^9}\cos(10\varphi).
 \label{app:eq:gl-tenth}
\end{equation}
At fixed nonzero $\rho$, minimization gives
\begin{equation}
 \begin{aligned}
 w<0:\quad &\cos(10\varphi)=1,
 &\varphi&=\frac{k\pi}{5},\\
 w>0:\quad &\cos(10\varphi)=-1,
 &\varphi&=\frac{\pi}{10}+\frac{k\pi}{5},
 \end{aligned}
 \qquad k\in\mathbb Z.
 \label{app:eq:gl-angle-minima}
\end{equation}
We next show explicitly that each set in Eq.~\eqref{app:eq:gl-angle-minima} forms a single symmetry orbit, and that the two sets are inequivalent under $D_5$ and time reversal.
We keep $\rho>0$ and the spin direction $\hat{\mathbf n}$ fixed when labeling configurations by $\varphi$ modulo $2\pi$.
From the transformation of $\boldsymbol\psi$ and Eq.~\eqref{app:eq:gl-collinear-angle},
\begin{equation}
 \begin{aligned}
 C_5:\quad &\varphi\mapsto\varphi-\frac{2\pi\ell}{5},\\
 \mathcal M:\quad &\varphi\mapsto-\varphi,\\
 \mathcal T:\quad &\varphi\mapsto\varphi+\pi.
 \end{aligned}
 \label{app:eq:gl-angle-actions}
\end{equation}
Here $\mathcal M$ is the reference spatial mirror, under which $\xi^{(1)}$ and $\xi^{(2)}$ are respectively even and odd. And $\mathcal{T}$ is time reversal
Thus a general angle in the symmetry orbit of $\varphi$ has the form
\begin{equation}
 \varphi'=\varepsilon\varphi-\frac{2\pi\ell n}{5}+t\pi
 \pmod{2\pi},
 \label{app:eq:gl-angle-orbit}
\end{equation}
where $\varepsilon=\pm1$, $n=0,\ldots,4$, and $t=0,1$.

For the first family, $\varphi_k^{\mathrm A}=k\pi/5$, a rotation $C_5^n$ brings the angle to $0$ modulo $\pi$ if
\begin{equation}
 k-2\ell n\equiv0\pmod5.
 \label{app:eq:gl-orbit-a}
\end{equation}
Since $\gcd(2\ell,5)=1$ for both $\ell=1$ and $\ell=2$, this congruence has a solution for every integer $k$.
The rotated texture is therefore either $\rho\xi_i^{(1)}\hat{\mathbf n}$ or its negative; time reversal removes the latter sign.
Consequently, every member of the first family is symmetry-equivalent to $\varphi=0$.

For the second family, $\varphi_k^{\mathrm B}=\pi/10+k\pi/5$, the corresponding condition for reaching $\pi/2$ modulo $\pi$ is
\begin{equation}
 \begin{gathered}
 \frac{\pi}{10}+\frac{(k-2\ell n)\pi}{5}
 \equiv\frac{\pi}{2}\pmod\pi,\\
 \Longleftrightarrow\quad k-2\ell n\equiv2\pmod5.
 \end{gathered}
 \label{app:eq:gl-orbit-b}
\end{equation}
The same coprimality ensures a solution for every $k$, with any remaining minus sign again removed by $\mathcal T$.
Thus the fact that $k=2$ yields $\varphi=\pi/2$ identifies a representative, while Eq.~\eqref{app:eq:gl-orbit-b} proves that all other members belong to its orbit.
In particular, $\varphi=\pi/10$ is mapped to $\pi/2$ by $C_5^4$ for $E_1$, or by $C_5^2$ for $E_2$, with angles understood modulo $2\pi$.

For $w<0$, choosing $k=0$ gives $\varphi=0$ and
\begin{equation}
 \mathbf m_i^{\mathrm A}
 =\rho\xi_i^{(1)}\hat{\mathbf n}.
 \label{app:eq:gl-pattern-a}
\end{equation}
For $w>0$, choosing $k=2$ gives $\varphi=\pi/2$ and
\begin{equation}
 \mathbf m_i^{\mathrm B}
 =\rho\xi_i^{(2)}\hat{\mathbf n}.
 \label{app:eq:gl-pattern-b}
\end{equation}
To distinguish the two orbits, consider $I(\varphi)=\cos(10\varphi)$.
Equation~\eqref{app:eq:gl-angle-orbit} gives
\begin{equation}
 \begin{aligned}
 I(\varphi')
 &=\cos\!\left(10\varepsilon\varphi-4\pi\ell n+10\pi t\right)\\
 &=\cos(10\varphi)=I(\varphi).
 \end{aligned}
 \label{app:eq:gl-orbit-invariant}
\end{equation}
On the two families, this invariant takes different values:
\begin{equation}
 \begin{aligned}
 I(\varphi_k^{\mathrm A})
 &=\cos(2k\pi)=+1=I(0),\\
 I(\varphi_k^{\mathrm B})
 &=\cos(\pi+2k\pi)=-1=I(\pi/2).
 \end{aligned}
 \label{app:eq:gl-two-orbits}
\end{equation}
In particular, $I(\pi/10)=I(\pi/2)=-1$, whereas $I(0)=+1$.
No combination of $D_5$ operations and time reversal can therefore transform the $\varphi=\pi/10$ texture into the pure $\xi^{(1)}$ texture.
Equivalently, such a transformation would require
\begin{equation}
 \varepsilon-4\ell n\equiv0\pmod{10},
 \label{app:eq:gl-orbit-obstruction}
\end{equation}
which is impossible because its left-hand side is odd.
Global spin rotations do not change this conclusion: for collinear textures they change $\hat{\mathbf n}$, and can reverse its sign, but do not continuously mix the two spatial modes.

Thus, A and B represent two inequivalent collinear families, with the sign of $w$ selecting which minimizes the leading angular anisotropy. For the coplanar state C, $q=0$ and this term vanishes. Higher-order angular terms may alter the
preferred mixing angles when the tenth-order term vanishes
or no longer dominates.

Within each family selected by the tenth-order anisotropy,
symmetry-related configurations have the same mean-field energy.
When the Hamiltonian and finite patch preserve the stated
symmetries, configurations within each family have the same
mean-field energy. It therefore suffices to compare their
representatives A and B, constructed from
$\boldsymbol{\xi}^{(1)}$ and $\boldsymbol{\xi}^{(2)}$, respectively.

\section{Mean-field energy comparison}
\label{app:mf_energy}
For each texture $p=\mathrm A,\mathrm B,\mathrm C$, the occupied mean-field
eigenstates define a zero-temperature Slater determinant $|\Phi_p\rangle$.
We evaluate the original Hubbard energy in this state. Define
\begin{equation}
 \mathcal H=H_t+U\sum_i\hat n_{i\uparrow}\hat n_{i\downarrow},\qquad
 H_t=-\sum_{ij,\sigma}t_{ij}c_{i\sigma}^\dagger c_{j\sigma}.
 \label{app:eq:physical-hubbard}
\end{equation}
The operator written in the main text is $H=\mathcal H-\mu\hat N_e$.
We first give the physical energy $E_p=\langle\Phi_p|\mathcal H|\Phi_p\rangle$.

Let $u_{\nu i}$ be the two-component spinor of eigenstate $\nu$ at site $i$.
Its occupation $f_{\nu p}\in\{0,1\}$ determines the one-body density matrix
\begin{equation}
 \begin{aligned}
 (P_{p,ij})_{\sigma\sigma'}
 &=\langle c_{j\sigma'}^\dagger c_{i\sigma}\rangle_p,\\
 P_{p,ij}&=\sum_\nu f_{\nu p}\,u_{\nu i}u_{\nu j}^\dagger.
 \end{aligned}
 \label{app:eq:density-matrix}
\end{equation}
At fixed integer $N_e$, occupy the lowest $N_e$ orbitals. Filling eigenvalues
below zero when $-\mu$ is included can instead yield different $N_e$ for
different textures. The output density and moment are
\begin{equation}
 \widetilde n_{p,i}=\operatorname{tr}_s P_{p,ii},\qquad
 \widetilde{\mathbf m}_{p,i}
       =\tfrac12\operatorname{tr}_s(\boldsymbol\sigma P_{p,ii}).
 \label{app:eq:output-moments}
\end{equation}
Here $\mathbf S_i=c_i^\dagger\boldsymbol\sigma c_i/2$ and
$c_i=(c_{i\uparrow},c_{i\downarrow})^{\mathsf T}$.
The output moments equal the input mean fields only at self-consistency.

For a Slater determinant without anomalous pairing, Wick's theorem gives
\begin{equation}
 \begin{aligned}
 \langle\hat n_{i\uparrow}\hat n_{i\downarrow}\rangle_p
 &=\langle\hat n_{i\uparrow}\rangle_p
   \langle\hat n_{i\downarrow}\rangle_p
   -|\langle c_{i\uparrow}^\dagger c_{i\downarrow}\rangle_p|^2\\
 &=\det P_{p,ii}
   =\frac{\widetilde n_{p,i}^{\,2}}{4}
       -|\widetilde{\mathbf m}_{p,i}|^2.
 \end{aligned}
 \label{app:eq:wick-double-occupancy}
\end{equation}
The spin-flip contraction must be retained for a coplanar state. Thus the
same expression applies to both collinear and noncollinear candidates:
\begin{equation}
 \boxed{E_p=\operatorname{Tr}(h_tP_p)
     +U\sum_i\left(\frac{\widetilde n_{p,i}^{\,2}}{4}
                         -|\widetilde{\mathbf m}_{p,i}|^2\right).}
 \label{app:eq:wick-energy}
\end{equation}
Here $(h_t)_{ij}=-t_{ij}\mathbb I_2$, $\mathbb I_2$ is $2\times2$ identity matrix; $\operatorname{tr}_s$ and
$\operatorname{Tr}$ trace spin and all one-body indices, respectively.
The energy per site is $e_p=E_p/N$. No extra double-counting correction is
added to Eq.~\eqref{app:eq:wick-energy}. It also holds for imposed textures
when $P_p$ and the moments are extracted from their occupied states.

For normalized spatial modes, input textures with the same total norm
$\sum_i|\mathbf m_i|^2=\rho^2$ can be parametrized as
\begin{equation}
 \begin{aligned}
 \mathbf m_i^{\mathrm A}&=\rho\xi_{i}^{(1)}\hat{\mathbf z},\qquad
 \mathbf m_i^{\mathrm B}=\rho\xi_{i}^{(2)}\hat{\mathbf z},\\
 \mathbf m_i^{\mathrm C}&=\frac{\rho}{\sqrt2}
             (\xi_{i}^{(1)}\hat{\mathbf x}+\xi_{i}^{(2)}\hat{\mathbf y}).
 \end{aligned}
 \label{app:eq:normalized-patterns}
\end{equation}
Fixed-amplitude comparisons and amplitude optimization are distinct.
At fixed filling, all candidates must have the same
$N_e=\operatorname{Tr}P_p$. For the operator $H$ used in the main text,
\begin{equation}
 \langle H\rangle_p=E_p-\mu N_{e,p}.
 \label{app:eq:chemical-convention}
\end{equation}
At common $N_e$ and $\mu$, this shift leaves the ordering unchanged;
otherwise the two energy conventions need not agree.
The preference for C is limited to the candidates and parameters examined.

The numerical comparison uses imposed trial textures rather than self-consistently converged magnetic states.
The spatial modes are orthonormal, and all three candidates have the same total squared moment, $\sum_i|\mathbf m_i|^2=m_0^2$, with $m_0=0.1$.
At each common chemical potential $\mu$, the occupied single-particle states determine $P_p$ and hence the Hubbard energy $E_p$ through Eq.~\eqref{app:eq:wick-energy}.
At zero temperature, the candidates are ranked by
\begin{equation}
 \Omega_p=E_p-\mu N_{e,p},
 \qquad N_{e,p}=\operatorname{Tr}P_p.
 \label{app:eq:comparison-grand-potential}
\end{equation}
The particle number may therefore differ between candidates.
We scan $U/t_0=1$--$15$ in steps of $0.5$ and $\mu/t_0=-2$--$0$ in steps of $0.1$, retaining points with a leading two-dimensional irreducible representation.
The amplitude is held fixed without optimization, and the resulting preference for C refers to this restricted comparison of trial states.
\section{Symmetry of the spin-resolved spectrum}
\label{sp:section}

\paragraph*{Spin projections and spectral weights.}
The single-particle Hamiltonian used to evaluate the observables is
\begin{equation}
 h=h_s\otimes\mathbb I_2
 +\sum_i|i\rangle\langle i|\otimes
 (\boldsymbol\Delta_i\cdot\boldsymbol\sigma),
 \qquad
 \boldsymbol\Delta_i=-\frac{2U}{3}\mathbf m_i,
 \label{app:eq:trial-exchange-field}
\end{equation}
where $\mathbf m_i$ is the imposed trial texture and $h_s$ contains the spin-independent hopping and scalar potentials. We choose spin-$z$ along the ordered
moments for collinear states and normal to the spin plane
for coplanar states. In the spin-$z$ basis,
\begin{equation}
 \ket{x\pm}=\frac{\ket{\uparrow}\pm\ket{\downarrow}}{\sqrt2},
 \qquad
 \ket{y\pm}=\frac{\ket{\uparrow}\pm i\ket{\downarrow}}{\sqrt2}.
 \label{sp:spin-basis}
\end{equation}
For a mean-field eigenstate $\ket{\lambda}$ with energy
$E_\lambda$, define
\begin{equation}
 a_{\lambda\sigma}(\mathbf k)
 =\braket{\mathbf k,\sigma|\lambda}
 =\frac{1}{\sqrt N}\sum_i e^{i\mathbf k\cdot\mathbf r_i}
       \braket{i,\sigma|\lambda}.
 \label{sp:spinor-amplitudes}
\end{equation}
The amplitudes along $x$ and $y$ are then
\begin{equation}
 a_{\lambda,x\pm}
 =\frac{a_{\lambda\uparrow}\pm a_{\lambda\downarrow}}{\sqrt2},
 \qquad
 a_{\lambda,y\pm}
 =\frac{a_{\lambda\uparrow}\mp i a_{\lambda\downarrow}}{\sqrt2}.
 \label{sp:projection-amplitudes}
\end{equation}
Writing $a_\lambda=(a_{\lambda\uparrow},
a_{\lambda\downarrow})^{\mathsf T}$ and
$w_\lambda=a_\lambda^\dagger a_\lambda$, the projected
spin expectation is
\begin{equation}
 \langle S^\nu\rangle_{\lambda,\mathbf k}
 =\frac{a_\lambda^\dagger\sigma_\nu a_\lambda}{2w_\lambda},
 \qquad w_\lambda>0.
 \label{sp:projected-spin}
\end{equation}

Using the Lorentzian
$L_\lambda(\omega)=
\eta/\{\pi[(\omega-E_\lambda)^2+\eta^2]\}$, define
\begin{equation}
 \begin{aligned}
 A_{\nu\pm}(\mathbf k,\omega)
 &=\sum_\lambda L_\lambda(\omega)
                  |a_{\lambda,\nu\pm}(\mathbf k)|^2,\\
 \mathcal A(\mathbf k,\omega)
 &=\sum_\lambda L_\lambda(\omega)
                  a_\lambda(\mathbf k)a_\lambda^\dagger(\mathbf k).
 \end{aligned}
 \label{sp:spectral-matrix}
\end{equation}
The spin splitting $D_\nu=A_{\nu+}-A_{\nu-}$ are
\begin{equation}
 \begin{aligned}
 D_x&=2\sum_\lambda L_\lambda
       \operatorname{Re}(a_{\lambda\uparrow}^*
                         a_{\lambda\downarrow}),\\
 D_y&=2\sum_\lambda L_\lambda
       \operatorname{Im}(a_{\lambda\uparrow}^*
                         a_{\lambda\downarrow}),\\
 D_z&=\sum_\lambda L_\lambda
       (|a_{\lambda\uparrow}|^2-|a_{\lambda\downarrow}|^2).
 \end{aligned}
 \label{sp:contrasts-explicit}
\end{equation}
Equivalently,
$\mathcal A=(A_0\mathbb I_2+\mathbf D\cdot\boldsymbol\sigma)/2$,
where $A_0=\operatorname{tr}_s\mathcal A$.
The normalized spectral spin expectation is
$\langle S^\nu\rangle_{\mathbf k,\omega}=D_\nu/(2A_0)$
for $A_0>0$.

Within the spin-group framework \cite{liuSpinGroupSymmetryMagnetic2022}, a symmetry combining a spatial operation $g$ with a spin rotation $U_g$ constrains the spectral matrix as
\begin{equation}
 \mathcal A(g\mathbf k,\omega)
 =U_g\mathcal A(\mathbf k,\omega)U_g^\dagger,
 \qquad
 \mathbf D(g\mathbf k,\omega)=R_g^s\mathbf D(\mathbf k,\omega),
 \label{sp:covariance}
\end{equation}
where $R_g^s$ is the corresponding spin-vector rotation.
The following relations hold at fixed $\omega$ and $\eta$.

\paragraph*{Collinear patterns A and B.}
For moments along $z$, the Hamiltonian is spin diagonal,
so $D_x=D_y=0$. Its real spin blocks also imply
$D_z(-\mathbf k)=D_z(\mathbf k)$.
Choose $\xi^{(1)}$ and $\xi^{(2)}$ to be even and odd,
respectively, under the reference mirror
$\mathcal M:(x,y)\mapsto(x,-y)$.
Pattern A preserves $\mathcal M$ alone, whereas pattern B
preserves $\mathcal M$ combined with
$U_x(\pi)=-i\sigma_x$. Thus,
\begin{equation}
 D_z^{\mathrm A}(\mathcal M\mathbf k)
 =D_z^{\mathrm A}(\mathbf k),
 \qquad
 D_z^{\mathrm B}(\mathcal M\mathbf k)
 =-D_z^{\mathrm B}(\mathbf k).
 \label{sp:collinear-mirror}
\end{equation}
Reflection therefore preserves each spin-resolved spectrum
for A but exchanges the opposite-spin spectra for B.
A fivefold rotation generally mixes the two spatial modes,
so a fixed A or B configuration need not have a
fivefold-invariant spectrum.
For collinear $A_1$ and $A_2$ states, the longitudinal
contrast is instead fivefold invariant, with even and odd
mirror parity, respectively.

\paragraph*{Coplanar pattern C.}
Following the main-text convention, take
\begin{equation}
 \mathbf m_i^{\mathrm C}
 =m_c\bigl(\xi_i^{(1)}\hat{\mathbf x}
          +\xi_i^{(2)}\hat{\mathbf y}\bigr).
 \label{sp:c-texture}
\end{equation}
For $R=C_5$ and $\theta=2\pi/5$, choose the doublet convention
\begin{equation}
 \begin{pmatrix}\xi^{(1)}(R\mathbf r)\\
                 \xi^{(2)}(R\mathbf r)\end{pmatrix}
 =\mathcal R(\ell\theta)
 \begin{pmatrix}\xi^{(1)}(\mathbf r)\\
                 \xi^{(2)}(\mathbf r)\end{pmatrix},
 \qquad \ell=1,2,
 \label{sp:mode-convention}
\end{equation}
where $\mathcal R$ is the two-dimensional rotation matrix
and $\ell$ labels $E_\ell$.
The equal-amplitude texture preserves the mirror combined
with $U_x(\pi)$ and the fivefold rotation combined with
$U_z(\ell\theta)=e^{-i\ell\theta\sigma_z/2}$.
For a $D_5$-symmetric scalar Hamiltonian,
Eq.~\eqref{sp:covariance} gives
\begin{equation}
 \begin{aligned}
 (D_x,D_y,D_z)(\mathcal M\mathbf k)
   &=(D_x,-D_y,-D_z)(\mathbf k),\\
 \begin{pmatrix}D_x\\D_y\end{pmatrix}(R\mathbf k)
   &=\mathcal R(\ell\theta)
       \begin{pmatrix}D_x\\D_y\end{pmatrix}(\mathbf k),\\
 D_z(R\mathbf k)&=D_z(\mathbf k).
 \end{aligned}
 \label{sp:coplanar-spatial}
\end{equation}
Thus, the $x$ contrast is mirror-even, while the $y$ and
$z$ splitting are mirror-odd. Fivefold rotations mix the
in-plane components but leave the normal component invariant.

Coplanarity also preserves the antiunitary operation
$\mathcal S=U_z(\pi)\mathcal T$, represented up to a phase
by $\sigma_x K$, where $K$ denotes complex conjugation. Consequently,
\begin{equation}
 \mathcal A(-\mathbf k)
 =\sigma_x\mathcal A(\mathbf k)^*\sigma_x,
 \qquad
 \begin{cases}
 D_{x,y}(-\mathbf k)=D_{x,y}(\mathbf k),\\
 D_z(-\mathbf k)=-D_z(\mathbf k).
 \end{cases}
 \label{sp:antiunitary}
\end{equation}
Hence, coplanarity does not require $D_z$ to vanish at
generic momentum, although the local out-of-plane
polarization is zero. Combining these constraints yields
\begin{equation}
 D_z(k,\phi,\omega)
 =\sum_{j\geq0}b_j(k,\omega)\sin[5(2j+1)\phi],
 \label{sp:z-harmonics}
\end{equation}
where $\phi$ is measured from the reference mirror.
The normal splitting therefore vanishes on the mirror
lines and has a fivefold-invariant signed pattern.

The same mirror and combined-rotation relations apply
to the spin-conductance contrasts when the contacts and
transport geometry are transformed consistently.
Thus, their angular dependence follows the same spatial
symmetry constraints as the corresponding spin-resolved angle-resolved photoemission spectroscopy (sARPES) splitting.
\section{Additional results for two-dimensional IRRPs}
\label{sm:more_results}
\begin{figure}[!htbp]
  \centering
  \begin{minipage}[t]{0.31\columnwidth}
    \vspace{0pt}
    \centering
    \begin{overpic}[width=\linewidth]
      {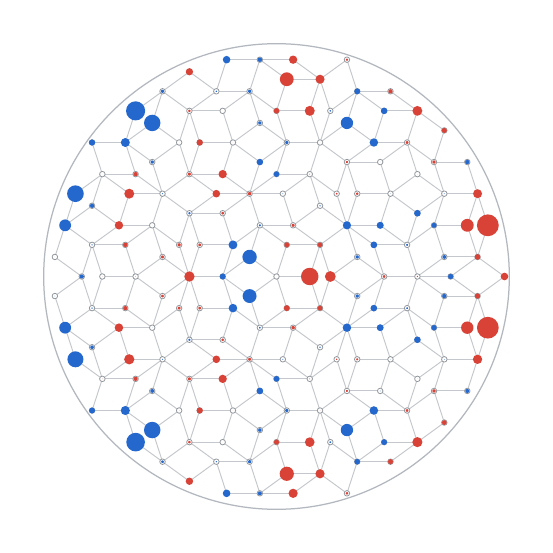}
      \put(3,94){\small\bfseries (a)}
    \end{overpic}
  \end{minipage}
  \hfill
  \begin{minipage}[t]{0.31\columnwidth}
    \vspace{0pt}
    \centering
    \begin{overpic}[width=\linewidth]
      {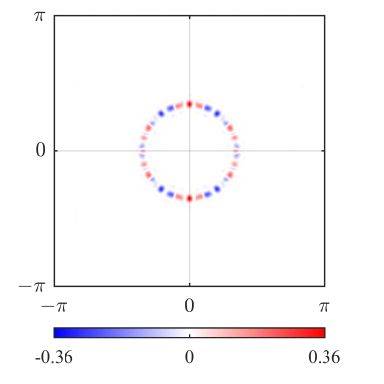}
      \put(15,90){\small
        \textbf{(b)} $\mA_{\uparrow}-\mA_{\downarrow}$}
    \end{overpic}
  \end{minipage}
  \hfill
  \begin{minipage}[t]{0.31\columnwidth}
    \vspace{0pt}
    \centering
    \begin{overpic}[width=\linewidth]
      {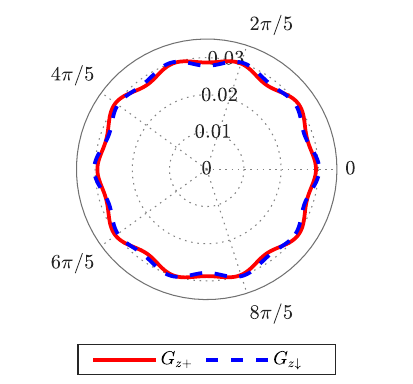}
      \put(3,80){\small\bfseries (c)}
    \end{overpic}
  \end{minipage}

  \par\vspace{0.5em}

  \begin{minipage}[t]{0.31\columnwidth}
    \vspace{0pt}
    \centering
    \begin{overpic}[width=\linewidth]
      {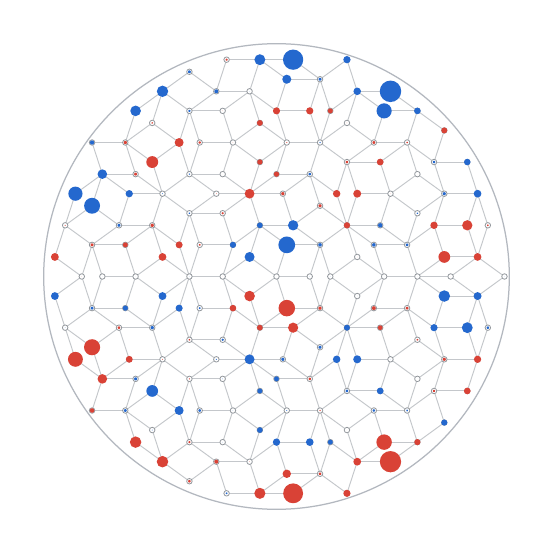}
      \put(3,94){\small\bfseries (d)}
    \end{overpic}
  \end{minipage}
  \hfill
  \begin{minipage}[t]{0.31\columnwidth}
    \vspace{0pt}
    \centering
    \begin{overpic}[width=\linewidth]
      {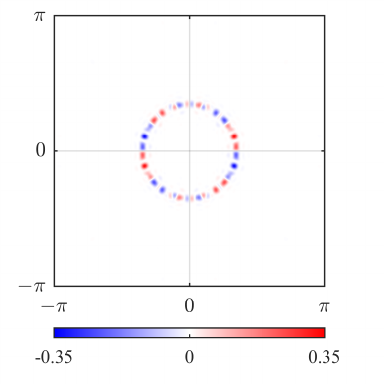}
      \put(15,90){\small
        \textbf{(e)} $\mA_{\uparrow}-\mA_{\downarrow}$}
    \end{overpic}
  \end{minipage}
  \hfill
  \begin{minipage}[t]{0.31\columnwidth}
    \vspace{0pt}
    \centering
    \begin{overpic}[width=\linewidth]
      {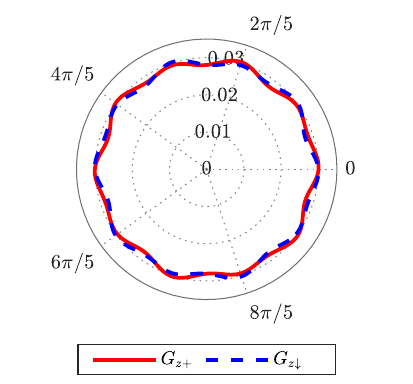}
      \put(3,80){\small\bfseries (f)}
    \end{overpic}
  \end{minipage}

  \caption{
    Collinear magnetic order in the $E_1$ IRRP and its
    observable signatures at $\mu/t_0=-1.6$.
    The upper and lower rows correspond to patterns A
    and B, respectively.
    (a,d) Real-space magnetic configurations, showing
    a central disk of radius $7a$ within the full
    radius-$28a$ Penrose patch.
    (b,e) Spin-resolved spectral differences at
    $\omega/t_0=-0.05$.
    (c,f) Angle-dependent spin conductance.
    Spin projections are resolved along $z$,
    parallel to the ordered moments.
    With the reference mirror $\mathcal{M}$ along
    the spatial $x$ axis, the magnetic configuration
    and spin splitting are mirror-even for A
    and mirror-odd for B.
  }
  \label{fig:collinear_m16}
\end{figure}

\begin{figure}[!htbp]
  \centering

  \begin{minipage}[t]{0.39\columnwidth}
    \vspace{0pt}
    \centering
    \begin{overpic}[width=\linewidth]
      {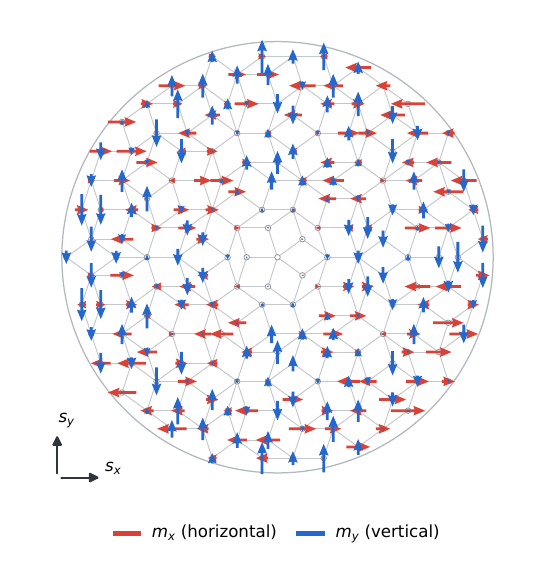}
      \put(3,90){\small\bfseries (a)}
    \end{overpic}
  \end{minipage}
  \hspace{0.06\columnwidth}
  \begin{minipage}[t]{0.39\columnwidth}
    \vspace{0pt}
    \centering
    \begin{overpic}[width=\linewidth]
      {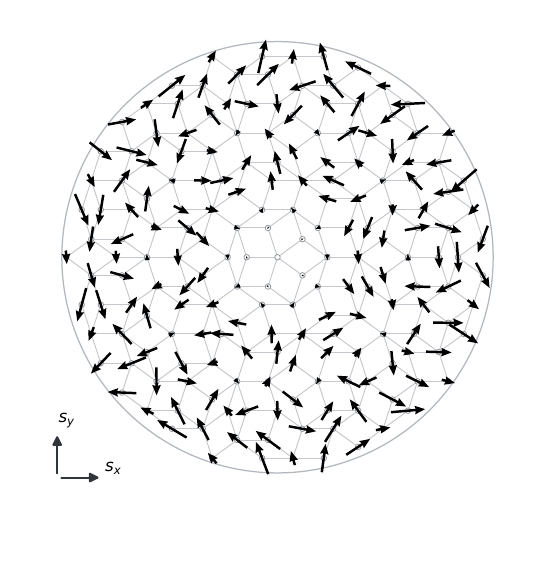}
      \put(3,90){\small\bfseries (b)}
    \end{overpic}
  \end{minipage}

  \par\vspace{0.5em}

  \begin{minipage}[t]{0.98\columnwidth}
    \centering
    \begin{overpic}[width=\linewidth]
      {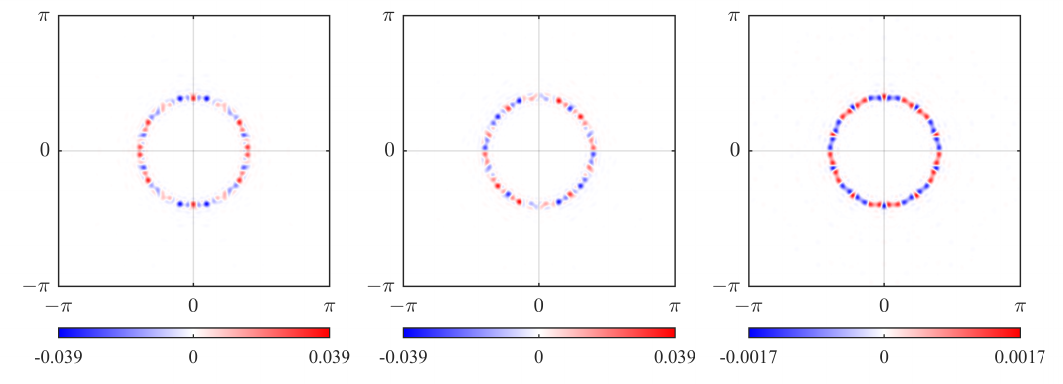}
      \put(6,32){\small
        \textbf{(c)} $\mA_{x+}-\mA_{x-}$}
      \put(38.5,32){\small
        \textbf{(d)} $\mA_{y+}-\mA_{y-}$}
      \put(71,32){\small
        \textbf{(e)} $\mA_{z+}-\mA_{z-}$}
    \end{overpic}
  \end{minipage}

  \par\vspace{0.5em}

  \begin{minipage}[t]{0.98\columnwidth}
    \centering
    \begin{overpic}[width=\linewidth]
      {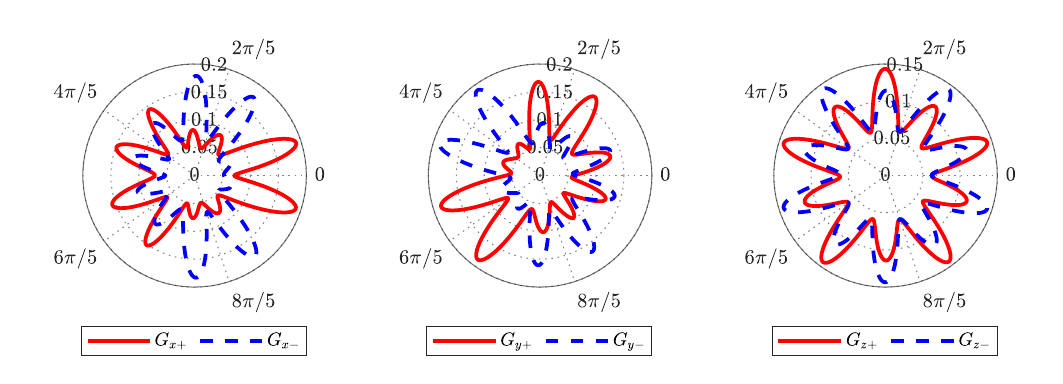}
      \put(3,32){\small \textbf{(f)} $x$}
      \put(35.5,32){\small \textbf{(g)} $y$}
      \put(68,32){\small \textbf{(h)} $z$}
    \end{overpic}
  \end{minipage}

  \caption{
    Coplanar magnetic order in $E_2$ IRRP and its observable signatures
    at $\mu/t_0=-1.1$.
    (a) Mirror-even $\boldsymbol{\xi}^{(1)}$
    (red, spin-$x$) and mirror-odd $\boldsymbol{\xi}^{(2)}$
    (blue, spin-$y$) components.
    (b) Their vector sum, forming coplanar pattern C;
    arrow directions and lengths indicate spin orientations
    and moment magnitudes, respectively.
    Panels (a,b) show a central disk of radius $7a$
    within the full radius-$28a$ Penrose patch.
    (c--e) Spin-resolved spectral differences along $x$, $y$,
    and $z$, respectively, at $\omega/t_0=-0.05$.
    (f--h) Corresponding angle-dependent conductance;
    red solid and blue dashed curves denote opposite
    spin projections.
    With the reference mirror along the spatial $x$ axis,
    the spin splitting is mirror-even in (c,f),
    mirror-odd in (d,g), and both mirror-odd and
    $C_5$-invariant in (e,h).
  }
  \label{fig:coplanar_m11}
\end{figure}
The main text presents collinear magnetic states in the
$E_2$ IRRP and a coplanar state in the $E_1$ IRRP.
Here we complement these results with collinear patterns
A and B in $E_1$ at $\mu/t_0=-1.6$ and coplanar pattern C
in $E_2$ at $\mu/t_0=-1.1$.
The spin axes, spectral definitions, and transport
conventions follow those of the main text.

For Figs.~S1 and S2, we use $U/t_0=5$ and $m_0=0.1$ with the normalized textures defined in Sec.~III.
The spectral maps are evaluated at $\omega/t_0=-0.05$ with Lorentzian broadening $\eta/t_0=0.005$.
The transport calculation uses two nonmagnetic semi-infinite chain probes at opposite in-plane positions $\pm R_{\mathrm{tip}}(\cos\phi,\sin\phi)$, with $R_{\mathrm{tip}}=2a$.
Each probe has nearest-neighbor hopping $-t_0$ and onsite energy $-1.2t_0$.
Its first five sites, separated vertically by $a$, couple to the sample through spin-independent hopping $-t_0e^{-d/a}$, where $d$ is the probe--sample distance.
Conductances are evaluated at the Fermi energy with retarded broadening $\eta_G/t_0=10^{-8}$, summing both incident spin channels and resolving the outgoing spin projection.

Figure~\ref{fig:collinear_m16} presents collinear patterns
A and B belonging to the $E_1$ doublet at $\mu/t_0=-1.6$.
Both are exactly compensated by orthogonality to the
uniform $A_1$ mode.
Pattern B additionally exhibits mirror-enforced pairwise
cancellation, whereas pattern A is mirror-even and has
no point-group operation that globally reverses its moments.
The same distinction appears in their sARPES and spin
conductance: reflection preserves each spin-resolved
response for A but exchanges opposite-spin responses
for B. The opposite-spin conductance curves nearly overlap because
their difference is small compared with the spin-independent
background. The spin splitting nevertheless follows the same
spatial symmetry constraints as the corresponding sARPES
splitting.
These results extend the collinear altermagnetic signatures discussed for $E_2$ in the main text to $E_1$.
The magnetic amplitude, interaction strength, spectral broadening, and probe geometry are specified at the beginning of Sec.~V.

Figure~\ref{fig:coplanar_m11} presents an additional coplanar
state belonging to $E_2$ at $\mu/t_0=-1.1$.
The mirror-even and mirror-odd modes combine along orthogonal
spin axes, forming the magnetic configuration in panel (b).
Its net magnetization vanishes because both modes are
orthogonal to the uniform $A_1$ mode.
The sARPES and spin conductance follow the same symmetry
relations as in the main text: reflection preserves the
opposite-spin responses separately for the $x$ projection
but exchanges them for the $y$ projection.
The $z$-resolved spin splitting is mirror-odd and invariant
under fivefold rotations.
The magnetic amplitude, interaction strength, spectral broadening, and probe geometry are specified at the beginning of Sec.~V.


\end{document}